\documentclass[twocolumn,preprintnumbers,amsmath,amssymb]{revtex4}

\usepackage{amsmath}    % need for subequations
\usepackage{graphicx}   % for figures
\usepackage{color}

\begin{document}

\title{Comment on ``The quantum mechanics of electric conduction in crystals,'' by R. J. Olsen and G. Vignale [Am. J. Phys. {\bf 78} (9), 954--960 (2010)]}
\author{Massimiliano Sassoli de Bianchi}
\affiliation{Laboratorio di Autoricerca di Base, 6914 Carona, Switzerland}\date{\today}
\email{autoricerca@gmail.com}   %optional

\begin{abstract}

In this note we use the notion of time-delay to explain the physical content of the transformation properties of transmission and reflection amplitudes, as a result of a displacement of the potential. Then, we reconsider the recent analysis of the scattering problem by a finite-periodic potential, by Olsen and Vignale, to obtain the total reflection condition in the limit of an infinite number of cells. In doing this, we obtain an expression of Hartman's effect, showing that the group velocity of the transmitted particle inside the potential chain can become arbitrary large, as the number of cells tends to infinity.


\end{abstract}

\maketitle

In a recent paper, R. J. Olsen and G. Vignale~\cite{Olsen} considered the one-dimensional scattering by a potential consisting of a chain of $N$ identical (non-overlapping) ``cells'', all separated by a same distance $a$. Their main goal was to study the limit $N\to\infty$ of a potential that becomes fully periodic on the positive half line. This problem is not new and has been studied by many authors, particularly in this journal; see for instance Refs.~\onlinecite{Griff, Blun, Sprun, New, Griff2} and the references cited therein.

To study the $N\to\infty$ limit, the authors of Ref.~\cite{Olsen} derive recurrence relations allowing them to express the transmission and reflection amplitudes of the finite-periodic potential in terms of those of its composing fragments. Their derivation is based on the tacit assumption that it is possible to construct the transmission (respectively, reflection) amplitudes by adding all the possible elementary scattering amplitudes that are associated to the different virtual paths the particle can follow inside the potential structure, before being ultimately transmitted (respectively, reflected). 

Let us observe that the validity of this assumption needs to be demonstrated. For this, one has to start from the Schroedinger equation and prove that, if the system is composed of two subsystems, the scattering matrix of the entire system factorizes, in the sense that its elements can be entirely expressed in terms of those associated to the single subsystems. This factorization property of the one-dimensional scattering matrix has been derived by a number of authors, with different methods (see our recent comment in this journal~\cite{Sas1}, and the references cited therein). Then, one observes that the obtained factorization formulae can be expressed as convergent power series, with each term of the series having the simple interpretation of describing a specific virtual path the particle can follow when it enters the chain, before being finally transmitted or reflected (see Ref.~\onlinecite{Sas2}, page 2730).

Having clarified the foundation of the approach used in Ref.~\onlinecite{Olsen}, we would like now to comment on some of the statements that were made in the article. In Sec. III, line 8 of the first paragraph, the authors affirm that: ``\textsl{Because the probability of an outcome is dependent only on the magnitude of the probability amplitude, there is no direct way to observe the phase of a particle.}'' This statement, as it is expressed, is not entirely correct. To give a counter example, the derivative with respect to energy of the phases of the transmission and reflection amplitudes are in principle directly observable, as they correspond, respectively, to the transmission and reflection time delays~\cite{Hau, Sas3}.

In Sec. III, line 4 of the third paragraph of Ref.~\onlinecite{Olsen}, one reads that: ``\textsl{The total transmission coefficient includes the geometric phase gained by an electron with wave vector $k$ as it travels through a distance $a$ [...].}'' We don't think that this statement is correct, as the transmission coefficient (or amplitude), contrary to the reflection ones, is not affected by the positioning of the potential, and consequently by the variation of the distance a particle has to travel to reach it. Rather than commenting the reasoning presented by the authors in Appendix B of Ref.~\onlinecite{Olsen}, in support of their assertion, let us show this fact explicitly. Let
\begin{equation}
\label{on-shell scattering matrix}
s =
\begin{pmatrix}
t \hfill & r \hfill \\
l \hfill & t \hfill
\end{pmatrix}
\end{equation}
be the scattering matrix of a one dimensional scattering system associated to a potential $V(x)$, with $t$ the transmission amplitude, and $l$ and $r$ the reflection amplitudes from the left and right, respectively, at energy $E = \hbar^2k^2/2m$ ($m$ being the mass of the particle). A rigid active translation of the potential to the right, through a distance $a$, i.e., $V(x)\to V(x-a)$, is implemented in quantum mechanics by the unitary displacement operator
\begin{equation}
\label{displacement operator}
D(a) = e^{\frac{i}{\hbar}pa}=
\begin{pmatrix}
e^{ika} \hfill & 0 \hfill \\
0 \hfill & e^{-ika} \hfill
\end{pmatrix},
\end{equation}  
where $p$ is the momentum operator. Therefore, the effect of the displacement on the $2\times 2$ scattering matrix is $s\to D^\dagger(a)sD(a)$, so that
\begin{align}
\label{displaced s-matrix}
s &\to
\begin{pmatrix}
e^{-ika} \hfill & 0 \hfill \\
0 \hfill & e^{ika} \hfill
\end{pmatrix}
\begin{pmatrix}
t \hfill & r \hfill \\
l \hfill & t \hfill
\end{pmatrix}
\begin{pmatrix}
e^{ika} \hfill & 0 \hfill \\
0 \hfill & e^{-ika} \hfill
\end{pmatrix}\nonumber\\
&= \begin{pmatrix}
t \hfill & re^{-2ika}\hfill \\
le^{2ika} \hfill & t \hfill
\end{pmatrix}.
\end{align}  
Eq. (\ref{displaced s-matrix}) clearly shows that the transmission amplitude is not affected by a rigid translation of the potential. Consequently, no additional phase factor is acquired by the transmission amplitude when the particle travels an extra distance to get to the interaction region. On the other hand, an additional phase factor is gained by the reflection amplitudes. Let us explain, by means of a simple physical argument, why it has to be so. 

As we mentioned above, the energy derivative of the phase $\alpha_t=\arg t$ of the transmission amplitude gives the transmission time-delay
\begin{equation}
\label{transmission time-delay}
\tau_t=\hbar\frac{d\alpha_t}{dE}=\frac{1}{v}\frac{d\alpha_t}{dk},
\end{equation}
where $v=\hbar k/m$ is the incoming velocity. Therefore, if it would be true that by moving the potential to a distance $a$ to the right, the transmission phase $\alpha_t$ would acquire an energy-dependent term $ka$, then, according to (\ref{transmission time-delay}), the transmission time-delay should be affected by the displacement and be increased by $a/v$. This however would be in contradiction with the very definition of time-delay, which is a relative quantity obtained by comparing an interaction time with a free reference time. When the interacting particle is transmitted, it necessarily travels the same distance as the free reference particle, independently of the position of the potential, which therefore cannot affect the transmission time-delay. Consequently, there cannot be any additional energy-dependent phase factor in the transmission amplitude. 

On the other hand, the situation is different for the reflection time-delays from the left and from the right, which are respectively given by
\begin{equation}
\label{reflection left time-delay}
\tau_l=\frac{1}{v}\frac{d\alpha_l}{dk}, \;\;\; \tau_r=\frac{1}{v}\frac{d\alpha_r}{dk},
\end{equation}
where $\alpha_l=\arg l$, and $\alpha_r=\arg r$. Indeed, contrary to a free reference particle, the total distance travelled by a reflected particle is affected by the positioning of the potential. If the particle comes from the left, and the potential is displaced by a distance $a$ to the right, it has to travel (forth and back) a supplementary distance $2a$, giving an additional free-flight contribution of $2a/v$ to the reflection time-delay from the left. After integration, this contribution yields the extra $2ka$ positive term in the phase of the reflection amplitude from the left, in agreement with (\ref{displaced s-matrix}). The situation for the reflection time-delay from the right is the opposite, as the displacement of the potential now decreases the total distance travelled by the reflected particle, in comparison to a free reference one, so that the reflection time-delay from the right is reduced of $2a/v$, giving this time a negative additional term $-2ka$ in the phase of the reflection amplitude from the right, also in agreement with (\ref{displaced s-matrix}).

In Sec. III of Ref.~\onlinecite{Olsen}, the authors derive recurrence formulae for the transmission and reflection amplitudes for a chain of equally spaced potential cells. The interpretation of their formulae (9) and (10) is of course partially invalidated by the above mentioned inexactness regarding the transformation properties of the transmission and reflection amplitudes. However, there is another reason for which we believe these expressions cannot be considered entirely correct. Although the authors are aware of the fact that for non-symmetric potentials the reflection amplitudes from the left and from the right differ (as they mention this in Sec. IV), they apparently overlooked this fact in their derivation. Indeed, even if we assume that the unit cell is symmetric (i.e., parity invariant), as soon as we add to it a second cell to the right, the symmetry of the two-potential system is lost. Therefore, in the derivation of the recurrence formulae for the transmission and reflection amplitudes, one needs to duly take into account the distinction between  reflection amplitudes from the left and from the right. Let us see how to do this. 

We introduce the following notation (which differs from the one adopted in Ref.~\onlinecite{Olsen}, for sake of clarity). We denote by $s_N$ the scattering matrix associated to the $N$-th single-cell, placed at a distance $(N-1)a$ from the first one, which is described by the scattering matrix $s_1\equiv s$. Also, we denote by $s^{(N)}$ the scattering matrix associated to the finite periodic chain of $N$ equally spaced cells, with $s^{(1)}=s_1=s$. Then, using the same notation for the elements of the scattering matrices, and considering that cells are iteratively added to the right of the chain, we have for the transmission amplitude:
\begin{align}
\label{transmission recurrence relation}
t^{(N+1)}&=t^{(N)}\left[1+l_{N+1}r^{(N)}+ \cdots\right]t_{N+1}\nonumber\\
&=\frac{t^{(N)}t_{N+1}}{1-l_{N+1}r^{(N)}}\\
&=\frac{t^{(N)}t}{1-lr^{(N)}e^{2ikNa}}, 
\end{align}
where the last equality follows from property (\ref{displaced s-matrix}). Similarly, for the reflection amplitude from the left, we have the recurrence relation:
\begin{align}
\label{reflection left recurrence relation}
l^{(N+1)}&=l^{(N)}+  t^{(N)}l_{N+1}\left[1+r^{(N)}l_{N+1}+ \cdots\right]t^{(N)}\nonumber\\
&=l^{(N)}+\frac{(t^{(N)})^2l_{N+1}}{1-l_{N+1}r^{(N)}}\\
&=l^{(N)}+\frac{(t^{(N)})^2le^{2ikNa}}{1-lr^{(N)}e^{2ikNa}}. 
\end{align}

Eqs. (\ref{transmission recurrence relation}) and (\ref{reflection left recurrence relation}) are what we believe to be the correct versions of the recursive statements (9) and (10) of Ref.\onlinecite{Olsen}, that duly take into account the transformation property (\ref{displaced s-matrix}) and the necessary distinction between the left and right reflection amplitudes. 

For later purpose, we also give the recurrence relation for the reflection amplitude from the right. In this case we have
\begin{align}
\label{reflection right recurrence relation}
r^{(N+1)}&=r_{N+1}+  t_{N+1}r^{(N)}\left[1+l_{N+1}r^{(N)}+ \cdots\right]t_{N+1}\nonumber\\
&=r_{N+1}+\frac{t_{N+1}^2r^{(N)}}{1-l_{N+1}r^{(N)}}\\
&=re^{2ik(N+1)a}+\frac{t^2r^{(N)}}{1-lr^{(N)}e^{2ikNa}}. 
\end{align}

It is also possible to derive a recurrence relation for the reflection amplitude from the right which is more similar to (\ref{reflection left recurrence relation}), in the sense of being also expressed in terms of the transmission amplitude $t^{(N)}$. For this, one has to proceed differently. Starting from a $N$-cells chain that is shifted to the right to a distance $a$, one adds a supplementary cell, but this time to the left, to obtain the $(N+1)$-cells structure. Then, duly taking into consideration the effects of the $a$-displacement on the phases of the reflection amplitudes, one finds
\begin{align}
\label{reflection right recurrence relation bis}
r^{(N+1)}&=r^{(N)}e^{-2ika}+  t^{(N)}r\left[1+l^{(N)}e^{2ika}r+ \cdots\right]t^{(N)}\nonumber\\
&=r^{(N)}e^{-2ika}+\frac{(t^{(N)})^2r}{1-l^{(N)}re^{2ika}}. 
\end{align}

Having established the recurrence relations for the scattering amplitudes, let us now try to reproduce the authors' argument, when they address the question of determining under which conditions the transmission amplitude $t^{(N)}$ goes to zero, as $N\to\infty$. From Eq. (\ref{reflection left recurrence relation}) it is clear that if we assume $t^{(N)}\to 0$, as $N\to\infty$, for a given $k$, then we must also have that $l^{(N+1)}= l^{(N)}+ o(1)$, in the same limit. This means that the reflection amplitude from the left tends to a constant as the number of cells increases, and by conservation of probability this constant has to be a pure phase factor, i.e., $l^{(\infty)} = e^{i\alpha_l^{(\infty)}}$. Olsen and Vignale use this fact to deduce that also the phase of the transmission amplitude $t^{(N)}$ tends to a constant as $N\to\infty$. To do so, they exploit the property that for a parity invariant potential the relative phase between $l^{(N)}$ and $t^{(N)}$ is $\pi/2$, i.e.,
\begin{equation}
\label{phases}
\alpha_l^{(N)}-\alpha_t^{(N)} = \pi/2\mod\pi. 
\end{equation}
However, as we already emphasized, the potential is not parity invariant and therefore one is not allowed to use (\ref{phases}) to infer that $\alpha_t^{(N)}$ also tends to a constant value, as $\alpha_l^{(N)}$ does. In fact, there is no way to infer such a result, as $\alpha_t^{(N)}$ \textit{does not} converge to a constant value, as $N\to\infty$. 

To see this, we start by observing that from the unitarity of the scattering matrix a general relation between the phases of the transmission and reflection amplitudes can be derived, also valid for non-symmetric potentials:
\begin{equation}
\label{phases general}
\frac{1}{2}\left(\alpha_l^{(N)}+\alpha_r^{(N)}\right)-\alpha_t^{(N)} = \pi/2 \mod\pi.
\end{equation}
By hypothesis, we know that $\alpha_l^{(N)}= \alpha_l^{(\infty)}+o(1)$. Thus, according to (\ref{phases general}), to determine the behavior of $\alpha_t^{(N)}$ for large $N$, we only need to determine that of $\alpha_r^{(N)}$. Using (\ref{reflection right recurrence relation bis}), we immediately find that, as $N\to\infty$,
\begin{equation}
\label{reflection right asymptotics}
r^{(N+1)}= r^{(N)}e^{-2ika} + o(1), 
\end{equation}
which means that, in this limit, 
\begin{equation}
\label{reflection phase right asymptotics}
\alpha_r^{(N+1)}=\alpha_r^{(N)}-2ka + o(1). 
\end{equation}
In other terms, the reflection amplitude from the right possesses the asymptotic form 
\begin{equation}
\label{reflection phase right asymptotics bis}
\alpha_r^{(N)}=\alpha - 2Nka + o(1), 
\end{equation}
where $\alpha$ is a $N$-independent (but energy-dependent) constant. It then follows from (\ref{phases general}) that the asymptotics of the transmission phase is 
\begin{equation}
\label{transmission phase right asymptotics}
\alpha_t^{(N)}=\beta - Nka + o(1),
\end{equation}
with $\beta$ a constant. Thus, we find that $\alpha_t^{(N)}$ diverges as $N\to\infty$. 

Let us open a parenthesis to illustrate the physical content of this result. According to (\ref{transmission phase right asymptotics}), the transmission time-delay for a $N$-cells finite-periodic potential and for a particle with incoming energy $E=\hbar^2k^2/2m = v^2/2m$, such that $t^{(N)}\to 0$, as $N\to\infty$, possesses the asymptotic behavior
\begin{align}
\label{Hartman}
\tau_t^{(N)}&=\frac{1}{v}\frac{d\alpha_t^{(N)}}{dk}=\frac{1}{v}\frac{d\alpha}{dk}-\frac{Na}{v} +o(1)\nonumber\\
&=-\frac{Na}{v}\left[1+o(1)\right].
\end{align}
It is interesting to observe that Eq. (\ref{Hartman}) is nothing but an expression of the so-called Hartman effect\cite{Har, Sas4}. Indeed, if we define the time $T^{(N)}_t$ spent by the transmitted particle inside the interaction region as the sum of the transmission time-delay $\tau_t^{(N)}$ and the time $Na/v$ it takes for a free particle of velocity $v$ to travel the distance $Na$, i.e.\cite{note},
\begin{equation}
\label{tunneling time}
T^{(N)}_t= \frac{Na}{v}+\tau_t^{(N)},
\end{equation}
then, inserting (\ref{Hartman}) into (\ref{tunneling time}), we find that $T^{(N)}_t\to 0$, as $N\to\infty$. This means that for sufficiently long finite-periodic chains, when the transmission probability tends to zero, the effective group velocity of the transmitted particle inside the potential becomes arbitrarily large, allowing for traversal group velocities even larger than the light speed in vacuum~\cite{note2}. 

We will now explain how to retrace the authors' argument to derive the condition for the total reflection, in the limit $N\to\infty$. For this, we multiply (\ref{transmission recurrence relation}) by $e^{ika}$ and write
\begin{equation}
\label{formula1}
\frac{t^{(N+1)}e^{ika}}{t^{(N)}}= \frac{te^{ika}}{1-lr^{(N)}e^{2ikNa}}.
\end{equation}
Because of (\ref{transmission phase right asymptotics}), the l.h.s. of (\ref{formula1}) tends to a real value as $N\to\infty$. Therefore, in the same limit, the imaginary part of the r.h.s. of (\ref{formula1}) must go to zero. Using (\ref{reflection phase right asymptotics bis}), we thus obtain the condition
\begin{equation}
\label{formula2}
\Im\frac{|t|}{e^{-i(\alpha_t+ka)}-|l|e^{i\theta}} = 0,
\end{equation}
where we have defined $\theta=\alpha + \alpha_l - \alpha_t - ka$. Since the numerator is real, the above condition is fulfilled if the imaginary part of the denominator is $0$, giving
\begin{equation}
\label{formula3}
\frac{\sin(\alpha_t+ka)}{|l|}= \sin\theta.
\end{equation}
This implies that
\begin{equation}
\label{formula4}
\frac{|\sin(\alpha_t+ka)|}{|l|}= |\sin\theta|\leq 1.
\end{equation}
Finally, using $|l|^2 = 1-|t|^2$, and $\sin^2x= 1 -\cos^2x$, we obtain that the condition for having total reflection in the limit of a potential that becomes fully periodic on the positive half line, is
\begin{equation}
\label{formula5}
\frac{\cos(\alpha_t+ka)}{|t|}\geq 1.
\end{equation}

We can observe that this condition is fully compatible with the one found in previous analysis by other authors~\cite{Griff, Sprun, Griff2}, where the Cayley-Hamilton theorem and Chebyshev polynomials of the second kind were used to solve the recurrence relations and express the transmission probability $|t^{(N)}|^2$ in terms only of the transmission amplitude of the (first) single cell:
\begin{equation}
\label{transmission probability}
|t^{(N)}|^2=\frac{1}{1+U_{N-1}^2(z)\frac{1-|t|^2}{|t|^2}},
\end{equation}
where $U_N(z)$ is the $N$-th Chebyshev polynomial of the second kind and $z=\cos(\alpha_t+ka)/|t|$. Now, by direct inspection of (\ref{transmission probability}), it is clear that $t^{(N)}\to 0$, as $N\to\infty$, iff $U_N(z)\to\infty$, in the same limit. Using the trigonometric identity
\begin{equation}
\label{trigonometric identity}
U_N(\cos\gamma)=\frac{\sin[(N+1)\gamma]}{\sin\gamma}
\end{equation}
and the properties of the hyperbolic functions, one can easily deduce that $|U_N(\cos\gamma)|\to\infty$, as $N\to\infty$, if $|\cos\gamma|>1$. This yields the total reflection condition $|z|=\cos(\alpha_t+ka)/|t|>1$, with a strict inequality, which is the one usually mentioned in the literature (see for instance Eq. (22) of Ref. \cite{Griff}). However, for the specific values $z=\pm 1$, we have for the Chebyshev polynomials the formulae: $U_N(1)=N+1$, and $U_N(-1)=(N+1)(-1)^N$. Therefore, it is also true that $|U_N(\pm 1)| = O(N+1)$, as $N\to\infty$, so that (\ref{transmission probability}) also gives condition (\ref{formula5}).

A final comment is in order. In Sec. III of Ref.~\onlinecite{Olsen} the authors affirm that: ``\textsl{There are also some conditions where the transmission coefficient of an infinite chain is nonzero [...]}.'' Strictly speaking, this statement is incorrect. In fact, although condition (\ref{formula5}) correctly describes the energies for which the transmission amplitude converges to zero, in the limit $N\to\infty$, one is not allowed to conclude that for energies breaking that condition, i.e., such that $\cos(\alpha_t+ka)/|t|<1$, the transmission probability would converge to a finite value, as $N\to\infty$. In fact, the limit of the transmission amplitude doesn't exist for those energies, as is clear from the fact that it becomes infinitely oscillating. However, if instead of a monoenergetic incoming wave one considers a wave packet whose energetic support is slightly smeared out, then, using the Riemann-Lebesgue Lemma and a power expansion, one can show that the transmission probability do converge in this case to a finite average value~\cite{New}.

In conclusion, in this note we have used the notion of time-delay to explain the physical content of the transformation properties of the transmission and reflection amplitudes, as a result of a displacement of the potential. We have then reconsidered the derivation in Ref.~\onlinecite{Olsen} to obtain the condition for total reflection, in the limit of an infinite number of cells composing the finite-periodic potential. In doing so, we have also obtained an expression of Hartman's effect, showing that the group velocity of the transmitted particle inside the chain can become arbitrary large, as $N\to\infty$.


\begin{thebibliography}{15}
\bibitem{note} Contrary to transmission and reflection time-delays, the notions of transmission and reflection sojourn times do not have a unique definition in quantum mechanics, as they correspond to the joint masurement of two incompatible observables. See references \cite{Hau, Sas3}, and those cited therein.
\end{thebibliography}
\end{document}